\documentclass[a4paper,11pt]{article}
\usepackage[latin1]{inputenc}
\usepackage{amssymb}

\title{{\bf   Decision  Problems For  Turing Machines }}

\author{Olivier Finkel   \\{\it   Equipe de Logique Math\'ematique}
  \\ CNRS et  Universit\'e Paris Diderot Paris 7
 \\ UFR de Math\'ematiques case 7012, site Chevaleret,\\75205 Paris Cedex 13, 
 France.\\ finkel@logique.jussieu.fr 
 \and 
Dominique Lecomte
\\{\it Institut de Math\'ematiques de Jussieu}\\ {\it  Projet Analyse Fonctionnelle}
\\ Universit\' e Paris 6, tour 46-0, bo\^\i te 186,
\\  4, place Jussieu, 75 252 Paris Cedex 05, France.
\\ dominique.lecomte@upmc.fr
\\and 
\\ Universit\'e de Picardie, I.U.T. de l'Oise, site de Creil,
\\ 13, all\'ee de la fa\"\i encerie, 60 107 Creil, France.}

\date{}
\begin{document}

\newtheorem{The}{Theorem}[section]
\newtheorem{Pro}[The]{Proposition}
\newtheorem{Deff}[The]{Definition}
\newtheorem{Lem}[The]{Lemma}
\newtheorem{Rem}[The]{Remark}
\newtheorem{Exa}[The]{Example}
\newtheorem{Cor}[The]{Corollary}
\newtheorem{Not}[The]{Notation}

\newcommand{\fa}{\forall}
\newcommand{\Ga}{\Gamma}
\newcommand{\Gas}{\Gamma^\star}
\newcommand{\Gao}{\Gamma^\omega}

\newcommand{\Si}{\mathsf\Sigma}
\newcommand{\Sis}{\Si^\star}
\newcommand{\Sio}{\Si^\omega}
\newcommand{\ra}{\rightarrow}
\newcommand{\hs}{\hspace{12mm}

\noi}
\newcommand{\lra}{\leftrightarrow}
\newcommand{\la}{language}
\newcommand{\ite}{\item}
\newcommand{\Lp}{L(\varphi)}
\newcommand{\abs}{\{a, b\}^\star}
\newcommand{\abcs}{\{a, b, c \}^\star}
\newcommand{\ol}{ $\omega$-language}
\newcommand{\orl}{ $\omega$-regular language}
\newcommand{\om}{\omega}
\newcommand{\nl}{\newline}
\newcommand{\noi}{\noindent}
\newcommand{\tla}{\twoheadleftarrow}
\newcommand{\de}{deterministic }
\newcommand{\proo}{\noi {\bf Proof.} }
\newcommand {\ep}{\hfill $\square$}
\renewcommand{\thefootnote}{\star{footnote}}

\maketitle 

\begin{abstract}
\noi  We  answer  two questions posed by Castro and Cucker in \cite{cc},  giving 
 the exact complexities of two decision problems about cardinalities of $\om$-languages of Turing machines.  
Firstly, it is $D_2(\Sigma_1^1)$-complete to determine whether the $\om$-language of a given Turing machine is countably infinite, where 
$D_2(\Sigma_1^1)$ is the class of $2$-differences of $\Sigma_1^1$-sets. 
Secondly, it is $\Sigma_1^1$-complete to determine whether the $\om$-language of a given Turing machine
is uncountable. 
\end{abstract}

\noi {\bf Keywords.} Theory of computation; computational complexity; formal languages; $\om$-languages; 
Turing machines; decision problems; analytical hierarchy. 

\section{Introduction}

Many classical decision problems arise  naturally in the fields of  Formal Language Theory and of Automata Theory. 
\nl Castro and Cucker studied decision problems for $\om$-languages of Turing machines in \cite{cc}. 
 Their motivation was, on the one hand, to classify problems about Turing machines and, on the other hand, to ``give natural complete problems for the 
lowest levels of the analytical hierarchy which constitute an analog of the classical complete problems given in recursion theory for the arithmetical hierarchy." 
They studied the degrees
 of many classical decision problems like : ``Is the $\om$-language recognized by a given machine non empty ?", ``Is it finite ?" ``Do two given machines 
recognize the same $\om$-language ?"
In particular, they proved that the non-emptiness and the infiniteness problems for $\om$-languages of Turing machines are $\Sigma_1^1$-complete, and that 
the universality problem, the inclusion problem, and the equivalence problem are $\Pi_2^1$-complete. Thus these problems are located at the first or the second level
of the analytical hierarchy and are ``highly undecidable." 
\nl Notice that Staiger studied also in \cite{Staiger93}  the verification property, which is in fact the inclusion problem, for many 
classes of $\om$-languages  located at one of the first  three levels of the arithmetical hierarchy. 
Cenzer and Remmel  studied in \cite{CenzerRemmel03} some decision problems for classes of $\om$-languages accepted by some computable deterministic 
automata. These classes of  $\om$-languages are located at one of the first three levels of the arithmetical hierarchy, 
while the class of $\om$-languages of Turing machines 
considered by Castro and Cucker and in this paper is actually the class of effective analytic sets. 
 Thus the class we consider in this paper is much larger than the classes studied by 
Cenzer and Remmel  in \cite{CenzerRemmel03}. Cenzer and Remmel studied also various {\it approximate} verification properties, determining the index sets for pairs of 
languages $(V, W)$ such that $W-V$ is finite, is a set of measure zero or contains only finitely many computable sequences. 
 The verification property is also studied by Klarlund in \cite{Klarlund94}. 
\nl  The following questions were left open by Castro and Cucker in \cite{cc}. What is the complexity of the following decision problems: 
``Is the $\om$-language recognized by a given Turing machine countably infinite ?", ``Is the $\om$-language recognized by a given Turing machine uncountable?"
\nl We answer here these questions, 
giving 
 the exact complexities of these two decision problems about cardinalities of $\om$-languages of Turing machines.  
Firstly, it is $D_2(\Sigma_1^1)$-complete to determine whether the $\om$-language of a given Turing machine is countably infinite, where 
$D_2(\Sigma_1^1)$ is the class of $2$-differences of $\Sigma_1^1$-sets. 
Secondly, it is $\Sigma_1^1$-complete to determine whether the $\om$-language of a given Turing machine
is uncountable. 
\nl This can be compared with this corresponding result of \cite{CenzerRemmel03,CenzerRemmel99}. It is $\Pi_3^0$-complete to determine 
whether a given $\Pi_1^0$ $\om$-language is infinite. 
It is $\Sigma^1_1$-complete to determine whether a given $\Pi_1^0$ $\om$-language is uncountable, and it is $\Pi_1^1$-complete to determine whether 
a given $\Pi_1^0$ $\om$-language is countably infinite, see \cite[Theroem 4.5]{CenzerRemmel99}. We refer the reader to \cite{CenzerRemmel03} for results about other 
classes of $\om$-languages, like the class of $\Sigma_1^0$ $\om$-languages, or the class of  $\Pi_2^0$ $\om$-languages. 

\section{Recall of basic notions}

\noi 
 The set of natural numbers is denoted by $\mathbb{N}$. 
We assume the reader to be familiar with the  arithmetical and  analytical hierarchies
on subsets of  $\mathbb{N}$, these notions  may be found in the textbooks on computability theory \cite{rog} 
\cite{Odifreddi1,Odifreddi2}. 
\nl We  now recall the notions of 1-reduction and of    $\Sigma^1_n$-completeness (respectively,           $\Pi^1_n$-completeness). 
Given two sets $A,B \subseteq \mathbb{N}$ we say $A$ is 1-reducible to $B$ and write $A \leq_1 B$
if there exists a total computable injective  function $f$ from      $\mathbb{N}$     to   $\mathbb{N}$        with $A = f ^{-1}[B]$. 
A set $A \subseteq \mathbb{N}$ is said to be $\Sigma^1_n$-complete   (respectively,   $\Pi^1_n$-complete)  iff $A$ is a  $\Sigma^1_n$-set 
 (respectively,   $\Pi^1_n$-set) and for each $\Sigma^1_n$-set  (respectively,   $\Pi^1_n$-set) $B \subseteq \mathbb{N}$ it holds that 
$B \leq_1 A$. 
An important fact is that, for each integer $n\geq 1$, there exist some $\Sigma^1_n$-complete subset of $\mathbb{N}$. 
Examples of such sets are  precisely described in \cite{rog} or \cite{cc}. 
In the sequel $E_1$ denotes  a  $\Sigma^1_1$-complete subset of $\mathbb{N}$. The set 
$E_1^-=\mathbb{N}-E_1  \subseteq \mathbb{N}$  is a $\Pi^1_1$-complete set.

\hs We assume now  the reader to be familiar with the theory of formal ($\om$)-languages  
\cite{Thomas90,Staiger97}.
We recall some  usual notations of formal language theory. 
\nl  When $\Si$ is a finite alphabet, a {\it non-empty finite word} over $\Si$ is any 
sequence $x=a_1\ldots a_k$, where $a_i\in\Si$ 
for $i=1,\ldots ,k$ , and  $k$ is an integer $\geq 1$.  
 $\Sis$  is the {\it set of finite words} (including the empty word) over $\Si$.
 \nl  The {\it first infinite ordinal} is $\om$.
 An $\om$-{\it word} over $\Si$ is an $\om$ -sequence $a_1 \ldots a_n \ldots$, where for all 
integers $ i\geq 1$, ~
$a_i \in\Si$.  When $\sigma$ is an $\om$-word over $\Si$, we write
 $\sigma =\sigma(1)\sigma(2)\ldots \sigma(n) \ldots $,  where for all $i$,~ $\sigma(i)\in \Si$. 
\nl 
 The usual concatenation product of two finite words $u$ and $v$ is 
denoted $u\cdot v$ and sometimes just $uv$. This product is extended to the product of a 
finite word $u$ and an $\om$-word $v$: the infinite word $u\cdot v$ is then the $\om$-word such that:
\nl $(u\cdot v)(k)=u(k)$  if $k\leq |u|$ , and 
 $(u\cdot v)(k)=v(k-|u|)$  if $k>|u|$.
\nl   
 The {\it set of } $\om$-{\it words} over  the alphabet $\Si$ is denoted by $\Si^\om$.
An  $\om$-{\it language} over an alphabet $\Si$ is a subset of  $\Si^\om$.

\hs Recall now  the notion of acceptance of infinite words by Turing machines considered  by Castro and Cucker in \cite{cc}. 

\begin{Deff}
A non deterministic Turing machine $\mathcal{M}$ is a $5$-tuple $\mathcal{M}=(Q, \Si, \Ga, \delta, q_0)$, where $Q$ is a finite set of states, 
$\Si$ is a finite input alphabet, $\Ga$ is a finite tape alphabet satisfying $\Si  \subseteq \Ga$, $q_0$ is the initial state, 
and $\delta$ is a mapping from $Q \times \Ga$ to subsets of $Q \times \Ga \times \{L, R, S\}$. A configuration of $\mathcal{M}$ is a triple 
$(q, \sigma, i)$, where $q\in Q$, $\sigma \in \Ga^\om$ and $i\in \mathbb{N}$. An infinite sequence of configurations $r=(q_i, \alpha_i, j_i)_{i\geq 1}$
is called a run of $\mathcal{M}$ on $w\in \Sio$ iff: 
\begin{enumerate}
\ite[(a)] $(q_1, \alpha_1, j_1)=(q_0, w, 1)$, and 
\ite[(b)] for each $i\geq 1$, $(q_i, \alpha_i, j_i) \vdash (q_{i+1}, \alpha_{i+1}, j_{i+1})$, 
\end{enumerate}
\noi where $\vdash$ is the transition relation of $\mathcal{M}$ defined as usual. The run $r$ is said to be complete if 
$(\fa n \geq 1) (\exists k \geq 1) (j_k \geq n)$. The run $r$ is said to be oscillating if $(\exists k \geq 1) (\fa n \geq 1) (\exists m \geq n) ( j_m=k)$. 

\end{Deff}

\begin{Deff}
Let $\mathcal{M}=(Q, \Si, \Ga, \delta, q_0)$ be a non deterministic Turing machine   and $F \subseteq Q$. The $\om$-language accepted by $(\mathcal{M}, F)$ is 
the set of $\om$-words $ \sigma \in \Sio$ such that there exists a complete non oscillating run $ r=(q_i, \alpha_i, j_i)_{i\geq 1}$
 of $\mathcal{M}$  on  $\sigma$ such that, for all $ i, q_i \in F.$
\end{Deff}

\noi The above acceptance condition is denoted $1'$-acceptance in  \cite{CG78b}. Other usual acceptance conditions are the now called B\"uchi or Muller
acceptance conditions, respectively denoted $2$-acceptance and $3$-acceptance in  \cite{CG78b}. 
 Cohen and Gold proved the following result  in \cite[Theorem 8.2]{CG78b}. 

\begin{The}[Cohen and Gold  \cite{CG78b}]
An $\om$-language is accepted by a non deterministic Turing machine with 
$1'$-acceptance condition iff it is accepted by a non deterministic Turing machine with B\"uchi (respectively, Muller) acceptance condition. 
\end{The}

\noi 
Notice that this result holds 
because  Cohen's and Gold's Turing  machines accept infinite words via {\it complete non oscillating runs}, while  $1'$, B\"uchi or Muller acceptance conditions
 refer to the sequence of states entered during an infinite run. 
\nl For other approaches, acceptance is based only on the sequence of states entered by the machine during an infinite computation \cite{Staiger97}, 
or one requires also that the machine reads the whole infinite tape \cite{eh}. 
We refer the reader to \cite{StaigerWagner78,Staiger99,StaigerFreund99,Staiger00} 
for a study  of these different approaches.

\hs We recall the existence of the arithmetical and analytical hierarchies  of $\om$-languages, see \cite{StaigerWagner78,Staiger97};  
see also \cite{LescowThomas} about logical specifications for infinite 
computations.  
The first class 
of the analytical  hierarchy is the class $\Sigma^1_1$ of {\it effective analytic sets} 
 which are obtained by projection of arithmetical sets. 
 By  \cite[Theorem 16]{Staiger99} (see also \cite[Theorem 5.2]{Staiger00})  
we have the following characterization of the class of   $\om$-languages accepted by non deterministic 
Turing machines via acceptance by complete runs (i.e., not necessarily non oscillating). 

\begin{The}[\cite{Staiger99}]\label{s}
The  class of  $\om$-languages 
accepted by non deterministic Turing machines with  $1'$ (respectively, B\"uchi,   Muller)  acceptance condition 
is  the class $\Sigma_1^1$ of  effective analytic sets.
\end{The}

\noi We return  now to Cohen's and Gold's non deterministic Turing machines accepting  via    {\it complete non oscillating runs}. 
The  following result follows from \cite[Note 2 page 12]{CG78b}  and from   Theorem \ref{s}. 

\begin{The}
The  class of  $\om$-languages 
accepted by Cohen's and Gold's non deterministic Turing machines with  $1'$ (respectively, B\"uchi,   Muller) acceptance condition 
is  the class $\Sigma_1^1$ of  effective analytic sets.
\end{The}

\section{Decision problems about  Turing machines}

\noi In the sequel we  consider, as in \cite{cc},  that the alphabet $\Si$ contains only two letters  $a$ and $b$, 
and we shall  denote $\mathcal{M}_z$ the non deterministic Turing machine of index $z$, reading words over $\Si$, 
equipped with a $1'$-acceptance condition.
 We now recall the results of  Castro and Cucker giving the exact complexity of the non-emptiness problem and of the infiniteness problem for 
$\om$-languages of Turing machines.

\begin{The}\label{U} 
\noi
\begin{enumerate}
\ite $\{ z \in \mathbb{N} \mid  L(\mathcal{M}_z) \neq \emptyset \}$ is $\Sigma_1^1$-complete. 
\ite $\{ z \in \mathbb{N} \mid  L(\mathcal{M}_z) \mbox{ is infinite} \}$ is $\Sigma_1^1$-complete. 
\end{enumerate}
\end{The}

\noi We now state our first new result. 

\begin{Lem}\label{2} 
$\{ z \in \mathbb{N} \mid  L(\mathcal{M}_z) \mbox{  is countably infinite} \}$ is in the class $D_2(\Sigma_1^1)$. 
\end{Lem}

\proo  We first show that $\{ z \in \mathbb{N} \mid  L(\mathcal{M}_z) \mbox{  is countable} \}$  is in the class $\Pi_1^1$. 
Notice that here ``countable" means ``finite or countably infinite."

\hs We know that an $\om$-language $L(\mathcal{M}_z)$  accepted by a Turing machine $\mathcal{M}_z$  is a $\Sigma_1^1$-subset of $\Sio$. 
But it is known that a $\Sigma_1^1$-subset $L$ of  $\Sio$    is  countable  if and only if for every $x\in L$ the singleton  $\{x\}$ is a $\Delta_1^1$-subset 
of $\Sio$, see \cite[page 243]{Moschovakis80}.

\hs On the other hand the following result is proved in \cite[Theorem 3.3.1]{HKL}. There esists a $\Pi_1^1$-set $W\subseteq \mathbb{N}$ and a 
$\Pi_1^1$-set $C \subseteq \mathbb{N}\times \Sio$ such that,  if we denote $C_n = \{x\in \Sio \mid (n, x) \in C \}$, then 
 $\{(n, \alpha) \in \mathbb{N}\times \Sio \mid  n\in W \mbox{ and } \alpha \notin C_n \}$ is a $\Pi_1^1$-subset 
of the product space $ \mathbb{N}\times \Sio$ and the $\Delta_1^1$-subsets of $\Sio$ are the sets of the form $C_n$ for $n\in W$. 

\hs 
We can  now   first  express $( \exists n \in W ~  C_n=\{x\} ) $  by the sentence $\phi(x)$:  
$$\exists n ~[ ~ n \in W   \mbox{ and }   (n,x) \in C \mbox{ and } \fa y \in \Sio ~ [ ( n\in W  \mbox{ and } (n,y) \notin C )  \mbox{ or } ( y=x ) ] ]$$

\noi But we know that $C$ is a $\Pi_1^1$-set and that  
$\{(n, \alpha) \in \mathbb{N}\times \Sio \mid  n\in W \mbox{ and } \alpha \notin C_n \}$ is a $\Pi_1^1$-subset 
of  $ \mathbb{N}\times \Sio$. Moreover the quantification  $\exists n $ in the above formula is a first-order quantification therefore the above formula 
$\phi(x)$ is a  $\Pi_1^1$-formula. 

\hs We can now  express that $L(\mathcal{M}_z) \mbox{  is countable}$ by the  sentence $\psi(z)$ : 
$$\fa x \in \Sio ~~ [ (  x \notin L(\mathcal{M}_z)  ) \mbox{ or } ( \exists n \in W ~  C_n=\{x\} ) ]$$
that is, 
$$\fa x \in \Sio ~~ [ (  x \notin L(\mathcal{M}_z)  ) \mbox{ or } \phi(x)  ]$$

\noi We know from \cite{cc} that $ x \notin L(\mathcal{M}_z) $ is expressed by a $\Pi_1^1$-formula. Thus the above formula $\psi(z)$
is a $\Pi_1^1$-formula. This proves that the set $\{ z \in \mathbb{N} \mid  L(\mathcal{M}_z) \mbox{  is countable} \}$  is in the class $\Pi_1^1$. 

\hs On the other hand 
the  set $\{ z \in \mathbb{N} \mid  L(\mathcal{M}_z) \mbox{ is infinite} \}$ is  in the class $\Sigma_1^1$. 
Finally the set $\{ z \in \mathbb{N} \mid  L(\mathcal{M}_z) \mbox{ is countably infinite} \}$ is the intersection of a $\Sigma_1^1$-set and of a 
$\Pi_1^1$-set, i.e. it is in the class $D_2(\Sigma_1^1)$. 
\ep

\hs We now give the exact complexity for this decision problem. 

\begin{The}\label{2} 
$\{ z \in \mathbb{N} \mid  L(\mathcal{M}_z) \mbox{  is countably infinite} \}$ is $D_2(\Sigma_1^1)$-complete. 
\end{The}

\proo  Recall that Castro and Cucker proved in \cite[Proof of Proposition 3.1]{cc} 
 that there is a computable injective function $\varphi$ from $\mathbb{N}$ into 
$\mathbb{N}$ such that there are two cases: 

\hs  {\bf First case:}  $z\in E_1$  and  $L(\mathcal{M}_{\varphi(z)})=\Sio$. 
\nl {\bf Second case:}   $z\in E_1^-$  and  $L(\mathcal{M}_{\varphi(z)})=\emptyset$. 

\hs We can easily define injective computable functions $g$ and $h$ from $\mathbb{N}$ into 
$\mathbb{N}$ suh that for every integer $z\in \mathbb{N}$ it holds that : 

$$L(\mathcal{M}_{g(z)})=L(\mathcal{M}_z) \cup a^\star\cdot b^\om$$
\noi and 
$$L(\mathcal{M}_{h(z)})=L(\mathcal{M}_z) \cap a^\star\cdot b^\om$$

\noi We can see that there are now two cases: 
\nl {\bf First case:} In this case  $z\in E_1$  and $L(\mathcal{M}_{\varphi(z)})=\Sio$. Thus $L(\mathcal{M}_{g\circ \varphi(z)})=\Sio$ is uncountable
and $L(\mathcal{M}_{h\circ \varphi(z)})=a^\star\cdot b^\om$ is countable. 
\nl {\bf Second case:} In this case  $z\in E_1^-$  and  $L(\mathcal{M}_{\varphi(z)})=\emptyset$.  
Thus $L(\mathcal{M}_{g\circ \varphi(z)})=a^\star\cdot b^\om$ is countable and $L(\mathcal{M}_{h\circ \varphi(z)})=\emptyset$.

\hs We define now the following simple operation over $\om$-languages. For two $\om$-words $x, x' \in \Sio$ the $\om$-word $x \oplus x'$ is defined by : 
for every integer $n\geq 1$ ~$(x \oplus x')(2n -1)=x(n)$ and $(x \oplus x')(2n)=x'(n)$. 
For two $\om$-languages $L, L' \subseteq \Sio$, the $\om$-language $L \oplus L' $ is defined by $L \oplus L' = \{ x \oplus x' \mid x\in L \mbox{ and } 
x'\in L' \}$. 

\hs It is easy to see that there is a computable injective function $\Phi$ from $\mathbb{N}^2$ into 
$\mathbb{N}$ suh that for every integer $z, z' \in \mathbb{N}$ it holds that 
$$L(\mathcal{M}_{\Phi(z,z')})=L(\mathcal{M}_z) \oplus L(\mathcal{M}_{z'}) $$

\hs We want now to show that every subset of  $\mathbb{N}$   in the class $D_2(\Sigma_1^1)$ is $1$-reducible to the set 
$\{ z \in \mathbb{N} \mid  L(\mathcal{M}_z) \mbox{  is countably infinite} \}$. 

\hs Let $E$ be a     $D_2(\Sigma_1^1)$-subset of   $\mathbb{N}$. Then there are   some sets  $A\subseteq \mathbb{N}$ and $B\subseteq  \mathbb{N}$
such that $A$ is a $\Sigma_1^1$-set and $B$ is a  $\Pi_1^1$-set and $E=A\cap B$.  But the set $E_1$ is $\Sigma_1^1$-complete and the set $E_1^-$ is 
$\Pi_1^1$-complete. Thus there are some injective computable mappings $f_A$ and $f_B$ from $\mathbb{N}$ into 
$\mathbb{N}$ suh that $A=f_A^{-1}(E_1)$ and $B=f_B^{-1}(E_1^-)$. 

\hs It is easy to see that there is an injective computable function $\Psi$ from $\mathbb{N}$ into 
$\mathbb{N}$ suh that for every $z\in \mathbb{N}$ it holds that :

$$L(\mathcal{M}_{\Psi(z)}) = L(\mathcal{M}_{\Phi(h\circ \varphi\circ f_A (z) , g\circ \varphi\circ f_B(z) )}) = 
L(\mathcal{M}_{h \circ \varphi\circ f_A (z) }) \oplus 
L(\mathcal{M}_{g \circ \varphi\circ f_B (z) })$$

\noi We next show that $\Psi$ is a reduction. We divide into cases.

\hs  {\bf First case:} $z\in E=A\cap B$. Then $f_A (z) \in E_1$ and $L(\mathcal{M}_{h\circ \varphi \circ f_A(z)})=a^\star\cdot b^\om$ is countably infinite. Moreover
 $f_B(z) \in E_1^-$  and $L(\mathcal{M}_{g\circ \varphi \circ f_B(z)})=a^\star\cdot b^\om$ is countably infinite. Thus the $\om$-language 
$L(\mathcal{M}_{\Psi(z)})=L(\mathcal{M}_{h \circ \varphi\circ f_A (z) }) \oplus 
L(\mathcal{M}_{g \circ \varphi\circ f_B (z) })$ is also  countably infinite. 
 
\hs  {\bf Second case:} $z\notin E=A\cap B$. Then either $z\notin A$ or  $z\notin B$.  Assume first that $z\notin A$. Then $f_A (z) \notin E_1$, i.e. 
$f_A (z) \in E_1^-$. Thus  $L(\mathcal{M}_{h\circ \varphi \circ f_A(z)})=\emptyset$. Assume now that $z\notin B$, i.e. 
$f_B (z) \in E_1$. Then   $L(\mathcal{M}_{g\circ \varphi \circ f_B(z)})=\Sio$.  We can see that if either $z\notin A$ or  $z\notin B$ the 
$\om$-language 
$L(\mathcal{M}_{\Psi(z)})=L(\mathcal{M}_{h \circ \varphi\circ f_A (z) }) \oplus 
L(\mathcal{M}_{g \circ \varphi\circ f_B (z) })$ can not be  countably infinite because it can only be either  empty or uncountable. 

\hs Finally, using the reduction $\Psi$ we have proved that 

$$E \leq_1 \{ z \in \mathbb{N} \mid  L(\mathcal{M}_z) \mbox{  is countably infinite} \}$$
\noi so this latter set is  $D_2(\Sigma_1^1)$-complete.
\ep

\begin{Rem}
Castro and Cucker noticed in \cite{cc} that the set $\{ z \in \mathbb{N} \mid  L(\mathcal{M}_z) \mbox{  is countably infinite} \}$ is in 
the class $\Sigma^1_2$ but they asked whether this set is $\Sigma^1_2$-complete. Our result shows that the answer is ``no" because a 
$D_2(\Sigma_1^1)$-set 
 is actually much less complex than a  $\Sigma^1_2$-complete set. 
\end{Rem}

\noi Recall that an $\om$-language accepted by a Turing machine $\mathcal{M}$  is a $\Sigma_1^1$-subset of $\Sio$. Then it is well 
known that such a set is either countable or has the cardinal $2^{\aleph_0}$ of the continuum, see  \cite{Moschovakis80}. Therefore an 
$\om$-language accepted by a Turing machine  has cardinal $2^{\aleph_0}$ iff it is not a countable set. We can now state the following result. 

\begin{The}\label{3} 
\noi
\begin{enumerate}
\ite $\{ z \in \mathbb{N} \mid  L(\mathcal{M}_z) \mbox{ is uncountable} \}$ is $\Sigma_1^1$-complete. 
\ite $\{ z \in \mathbb{N} \mid  L(\mathcal{M}_z) \mbox{ is  countable} \}$ is $\Pi_1^1$-complete. 
\end{enumerate}
\end{The}

\proo We first prove item $(1)$. 
\nl  We have already seen that $\{ z \in \mathbb{N} \mid  L(\mathcal{M}_z) \mbox{ is countable} \}$ is in the class $\Pi_1^1$. Thus the set 
$\{ z \in \mathbb{N} \mid  L(\mathcal{M}_z) \mbox{ is uncountable} \}$ is a $\Sigma_1^1$-set. 

\hs To prove the completeness result we can use an already cited result of Castro and Cucker. 
 There is a computable injective function $\varphi$ from $\mathbb{N}$ into 
$\mathbb{N}$ for which one of the two following cases hold: 
\nl {\bf First case:} $z\in E_1$  and  $L(\mathcal{M}_{\varphi(z)})=\Sio$. 
\nl {\bf Second case:} $z\in E_1^-$  and  $L(\mathcal{M}_{\varphi(z)})=\emptyset$. 

\hs The reduction $\varphi$ shows that :
$$E_1 \leq_1  \{ z \in \mathbb{N} \mid  L(\mathcal{M}_z) \mbox{ is uncountable} \}$$
\noi so this latter set is $\Sigma_1^1$-complete. 

\hs Item $(2)$ follows directly from Item $(1)$. 
\ep

\end{document}